%% file: main.tex
\documentclass[a4paper, amsfonts, amssymb, amsmath, reprint, showkeys, nofootinbib, twoside]{revtex4-1}
\usepackage[english]{babel}
\usepackage[utf8]{inputenc}
\usepackage[colorinlistoftodos, color=green!40, prependcaption]{todonotes}
\input{preamble}
\usepackage[pdftex, pdftitle={Article}, pdfauthor={Author}]{hyperref} 
\begin{document}

\title{Generation of Tunable Three-Photon Entanglement in Cubic Nonlinear Coupled Waveguides}

\author{Miguel Y. Bacaoco$^{1,4}$}
\author{Max J. Galettis$^2$}
\author{James Huang$^3$}
\author{Denis Ilin$^{1,4}$}
\author{Alexander S. Solntsev$^{1,4}$}
    \email[Correspondence email address: ]{alexander.solntsev@uts.edu.au}
    \affiliation{$^1$University of Technology Sydney, NSW 2007, Australia \\ 
    $^2$University of New South Wales, NSW 2052, Australia \\
    $^3$The University of Sydney, NSW 2006, Australia \\
    $^4$Sydney Quantum Academy, NSW 2006, Australia \\
    }


\begin{abstract}
We theoretically investigate the generation of three-photon states with spatial entanglement in cubic nonlinear coupled waveguides using third-order spontaneous parametric down-conversion and quantum walks. Our approach involves independently pumping two coupled waveguides to generate a path-encoded three-photon Greenberger–Horne–Zeilinger (GHZ) state, which then evolves with complex spatial dynamics governed by coupling coefficients and phase mismatch. By appropriate parameter tuning, we demonstrate the generation of robust heralded Bell states, uniform states, and GHZ-like states at the chip output. This work demonstrates an integrated source of three-photon spatial entanglement on a simple chip, offering additional reconfigurability for advanced multiphoton quantum applications.
\end{abstract}

\keywords{Entanglement, Quantum Photonics, Nonlinear Optics, Integrated Photonics}

\maketitle

\section{Introduction}

Integrated photonic quantum state sources are attractive for miniaturizing and scaling emerging quantum technologies. \cite{Wang2020,Giordani2023_IPinQT,Ramakrishnan2023_IPPforQT,Labonte2024_IPQCommMet} This is due to the mature fabrication industry and their ease of operation without requiring millikelvin temperatures or high vacuum conditions. ~\cite{Wang2020} Recent demonstration of chip-scale quantum key distribution (QKD),~\cite{Sibson2017,Wang2016} universal linear-optical circuits,~\cite{Carolan2015} and quantum teleportation,~\cite{Metcalf2014} highlights the potential of integrated photonics to implement key quantum technologies on a scalable platform. Furthermore, hybrid integration of quantum photonic components with semiconductor optoelectronic devices, such as single-photon emitters, detectors, lasers, and integrated circuits, enables efficient and compact quantum information processing, sensing, and communication on a chip. \cite{Wang2022,Elshaari2020,Soref2023}

Entangled photons are of particular importance in quantum teleportation and cryptography, and have been routinely realized in both bulk and integrated optics using single-crystal beta barium oxide (BBO), ~\cite{Bedington2017} or periodically-poled nonlinear materials such as lithium niobate (LiNbO3),~\cite{Tanzilli2002,Jin2014} and potassium titanyl phospate (KTP).~\cite{Bedington2017} These sources emit polarization-entangled photon pairs via Type-II quadratic $\chi^{(2)}$ spontaneous parametric down-conversion (SPDC).~\cite{Couteau2018} The manipulation of the generated entangled state then relies on succeeding optical components such as beam splitters, electro-optic phase-shifters and modulators, polarizers, etc. with application-specific properties for post-processing. Simultaneous generation and manipulation of entangled photonic states on integrated devices offer a compact, scalable, and efficient platform due to its reduced form factor and resilience to environment-induced decoherence. Such devices have been recently realized on a nonlinear waveguide arrays with optical switching able to generate Bell states and N00N states on demand.~\cite{Solntsev2012_PRL, Solntsev2012_cubicSFWM, Solntsev2014, Setzpfandt2016,Solntsev2017} The scheme utilized nonlinear material as the waveguide core which generates photon-pairs via SPDC upon optical excitation. The  coupling between the waveguides allows tunable wave-mixing, depending only on the optical properties of the pump, which results to the desired quantum state at the waveguide output.

Interest in tripartite entanglement beyond the typical qubit requires alternative sources of entangled photons and has pushed for the realization of third-order SPDC (TOSPDC) in cubic $\chi^{(3)}$ nonlinear materials that produce the photon-triplet state. ~\cite{Bencheikh2007,  Chekhova2005, Okoth2019,Cavanna2020, Banic2022} Among the sought-after properties unique to photon-triplets is its super-Gaussian quantum statistics that is of particular importance in continuous-variable quantum information.~\cite{Agusti2020, Zhang2021} Another application is in heralding optical Bell states, which need at least one additional quantum state to herald and optical Bell state.~\cite{Hamel2014} Hence, generating and manipulating photon triplets into valuable states for various quantum applications has become a subject of fundamental and practical research.

In this work, we theoretically introduce a method to generate and manipulate three-photon correlations in a nonlinear directional coupler without the need for post-processing. Our approach uses three-photon quantum states produced via third-order spontaneous parametric down-conversion (TOSPDC) in a cubic nonlinear waveguide, which propagate with quantum walk, providing inherent post-processing for the generated photons. At the waveguide exit, our system exhibits spatial three-photon correlations and entanglement, tunable via the phase mismatch and coupling coefficients of the photon modes --- able to produce heralded Bell states, uniform states, and GHZ-like states.

\section{Concept and Theoretical Model}

\begin{figure}
     \centering
     \includegraphics[width=\linewidth]{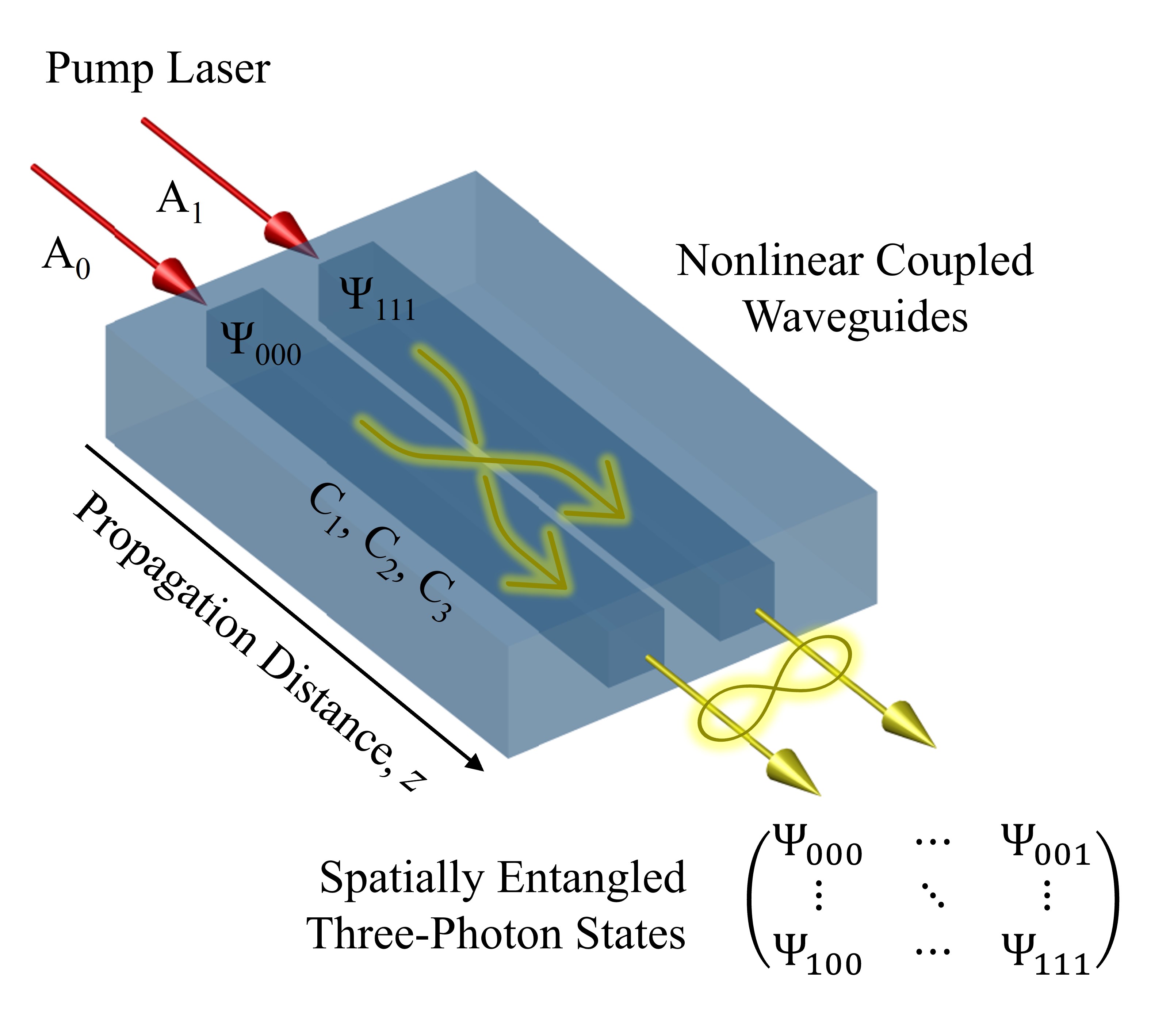}
     \caption{Schematic representation of two nonlinear coupled waveguides independently excited by coherent pumps $A_0$ and $A_1$ producing the photon triplet states $\Psi_{000}$ and $\Psi_{111}$, respectively, via third-order SPDC. The generated states propagate and couple to the other waveguide with coupling coefficient $C_1$, $C_2$, and $C_3$, for each photon mode, which results to tunable output quantum state $\Psi_{l,m,n}$ where $l,m,n \in \{0,1\}$ at the waveguide exit.}
     \label{fig:schematic}
 \end{figure}

We consider a directional coupler composed of two cubic nonlinear waveguides suspended on a semi-infinite cladding schematically shown in Figure \ref{fig:schematic}. Each waveguide is excited by a coherent linearly-polarized pump of frequency $\omega_p$ and undergoes TOSPDC producing three quantum-correlated photons with frequencies $\omega_1$, $\omega_2$, and $\omega_3$ (where $\omega_p = \omega_1 + \omega_2 + \omega_3$), which then constitute the photon triplet state $|\Psi(\omega_1, \omega_2, \omega_3)\rangle$. 

The phase mismatch for TOSPDC is given by $\Delta\beta \equiv \beta_p - \beta_1 -\beta_2 - \beta_3$, where $\beta_i$ is the propagation constant of the $i$th photon's transverse mode along the z-axis. Here, we defined the phase mismatch parameter $\Delta \beta$ that determines the efficiency of the nonlinear process according to $\varphi(\Delta \beta, z) \equiv \text{sinc}(\Delta \beta z /2)\exp{(-i \Delta \beta z/2)}$. ~\cite{Boyd2008} Additionally, due to the local coupling of the transverse modes with wavenumber $k_i^\perp$ between the waveguides, the propagation constant disperses in accordance to discrete diffraction in infinite waveguide arrays, which effectively takes the form: ~\cite{Lederer2008,Solntsev2012_PRL} 

\begin{equation}
   \beta_i = \beta_i(\omega_i,k_i^\perp) = \tilde{\beta_i}(\omega_i) + 2C_i \cos{(k_i^\perp)}.
   \label{eqn:dispersion}
\end{equation}

The propagation constant follows this dispersion relation in terms of the coupling coefficient $C_i$ that characterizes the strength of the modal overlap and, consequently, the quantum walk across the two waveguides due to photon tunneling.~\cite{Solntsev2012_PRL} At the pump frequency, the same dispersion applies, however, due to weak mode overlap at higher frequencies,~\cite{Lederer2008} such coupling is generally much smaller than the generated photons i.e. $C_p \ll  C_{\{1,2,3\}}$, and thus negligible. This can be more intuitively understood in terms of the waveguide mode of the pump and the generated photons where the latter, having approximately three times the wavelength of the pump, will have much larger spatial guided profile than that of the pump. For the TOSPDC photons to start having substantial modal overlap, the pump photons are already separated by a considerable distance that limit their interaction. By setting $C_p=0$, we neglect such coupling effects and assume that the pump beam profile remains undepleted in a single waveguide along the propagation distance. 

The overall phase mismatch may then be expressed in terms of the individual $\beta_i(\omega_i, k_i^\perp)$ as:

\begin{equation}
    \Delta \beta =  \tilde{\beta}(\omega_p) - \sum_{i=1}^{n=3} \left[\tilde{\beta}_i(\omega_i) + 2C_{i} \cos{ (k_i^\perp)}\right].
    \label{eqn:phase_mismatch}
\end{equation}

Upon optical excitation at the waveguide input, the coherent pump facilitates third-order SPDC producing pure photon-triplet quantum states of the form: ~\cite{Bencheikh2022,Banic2022} 

\begin{align}
\begin{split}
|\psi_{k}\rangle = B_0 \sum_{k_1,k_2,k_3}  \int d\omega_1 d\omega_2 d\omega_3 | \Psi_k(\omega_1, \omega_2, \omega_3,z) \rangle,
\end{split}
\end{align}

where
\begin{align}
\begin{split}
|\Psi_{k}(&\omega_1, \omega_2, \omega_3,z)\rangle = \Omega(k_p, k_1, k_2, k_3) \varphi(\Delta \beta, z)\\ 
&\times \hat{a}^\dag(\omega_1) \hat{a}^\dag(\omega_2) \hat{a}^\dag(\omega_3) |0,0,0\rangle.
\end{split}    
\end{align}

Here, $B_0$ is a constant, $\hat{a}^\dag$ is the usual bosonic creation operator, and $\Psi_k$ is the three-photon wavefunction in momentum space. The term $\Omega$ describes the overlap of the pump beam and the transverse Bloch modes defined as $\Omega(k_p, k_1, k_2, k_3) = \sum_m \gamma A_k (k_p)\exp{[i(k_p-k_1-k_2-k_3)m]}$, where $A_k$ is the k-space representation of the input pump $A_p(n')$ at waveguide $n'$. and $\gamma$ is the third-order nonlinear strength. \cite{Grafe2012} By performing Fourier transform to the three-photon wavefunction in momentum space $|\Psi_k \rangle \rightarrow |\Psi_{l,m,n}\rangle$, we get its real-space representation with spatial basis  $l,m,n$ for the three respective photon modes $\omega_1$, $\omega_2$, and $\omega_3$. Here, the spatial basis $l, m, n$ can be projected as the indices of a one-dimensional array of coupled waveguides. Following similar mathematical argument in Reference~\cite{Grafe2012}, the spatial evolution of the three-photon wavefunction in a coupled waveguide array can be described by the following coupled-mode differential equation:

\begin{align}
\begin{split}
-i\frac{d}{dz}\Psi_{l,m,n} &= C_1 \left[ \Psi_{l-1,m,n} + \Psi_{l+1,m,n} \right]\\ 
&+ C_2 \left[ \Psi_{l,m-1,n} + \Psi_{l,m+1,n} \right] \\ 
&+ C_3 \left[ \Psi_{l,m,n-1} + \Psi_{l,m,n+1} \right]  \\&- i \sum_{p=0,1}  \delta_{m,p} \delta_{n,p} \delta_{l,p}  \gamma A_{p}  e^{i\Delta \beta z}.
\end{split}
\label{eqn:spatial_evolution}
\end{align}

For the directional coupler considered in this study, we truncate the spatial basis to only two waveguides with indices $0$ and $1$ i.e. $l,m,n \in \{0,1\} $. This leads to solving the solution $\Psi_{l,m,n}$ of eight coupled differential equations.

In the trivial case where the waveguides are uncoupled i.e. $C_i =0$, the resulting wavefunction resembles that of a three-photon Greenberger–Horne–Zeilinger (GHZ) state, $|\text{GHZ}\rangle = (|\Psi_{000}\rangle+|\Psi_{111}\rangle)/\sqrt{2}$. This state itself is already useful for fundamental tests of quantum mechanics and quantum information~\cite{Greenberger1989,Bouwmeester1999} that, as we have shown, can be prepared without post-selection in our proposed system in contrast to other schemes.~\cite{Hamel2014, Ju2019} In this work, the nontrivial solution to the differential equation was investigated to seek other three-photon correlations and entanglement that are of practical relevance to advanced and emerging integrated photonics quantum technologies.

\section{Results and Discussion}

Analytic examination of Equation~\ref{eqn:spatial_evolution} reveals that there are at most four linearly independent eigenvalues (see Supporting Information). By equalizing and normalizing the pump amplitudes $A_0$ and $A_1$, the system achieves symmetry with respect to the permutation of waveguides. This symmetry simplifies the analysis and enhances the tunability of the quantum states generated, facilitating the controlled production of various entangled states. The analytic solution of the wavefunction results in:

\begin{equation}
\begin{aligned}\label{eq:sol_gen}
    &\Psi_{000}=\Psi_{111}=\frac{1}{2\sqrt{2}}(a+b+c+d), \\ 
    &\Psi_{100}=\Psi_{011}=\frac{1}{2\sqrt{2}}(a-b-c+d), \\
    &\Psi_{010}=\Psi_{101}=\frac{1}{2\sqrt{2}}(-a+b-c+d), \\
    &\Psi_{001}=\Psi_{110}=\frac{1}{2\sqrt{2}}(-a-b+c+d).
\end{aligned}
\end{equation}

Here, $a,b,c,d=\Phi_a(z),\Phi_b(z),\Phi_c(z),\Phi_d(z)$ are the  ansatz to the wavefunction of form $\Phi_j(z) \propto -[1-\exp{(i \lambda_j z)}]/i\lambda_j$ corresponding to  eigenvalues $\lambda_j$ defined as:

\begin{equation}
\begin{aligned}\label{eq:eigenvalues}
    &\lambda_a = -\Delta \beta+C_1-C_2-C_3, \\ 
    &\lambda_b = -\Delta \beta-C_1+C_2-C_3, \\
    &\lambda_c = -\Delta \beta-C_1-C_2+C_3, \\
    &\lambda_d = -\Delta \beta+C_1+C_2+C_3.
\end{aligned}
\end{equation}

Clearly, it can be seen that the overall wavefunction is simply a linear combination of four independent oscillatory and linear functions over $z$, whose properties depend only on optically controllable system parameters $\Delta \beta$ and $C_{\{1,2,3\}}$. We also note that if $\lambda_j=0$, the oscillatory term vanishes and $\Phi_j$ simply becomes linear i.e. $\Phi_j(z) \propto z$. In the succeeding sections, we confine our study on waveguides with length $L=z_{\max}=2\pi$, coupling coefficients $C_1=C_2,C_3>0$, and $\gamma=1$, independently pumped by a coherent source $A_0=A_1=1$. 

\begin{figure}
\centering
\includegraphics[width=0.95\linewidth]{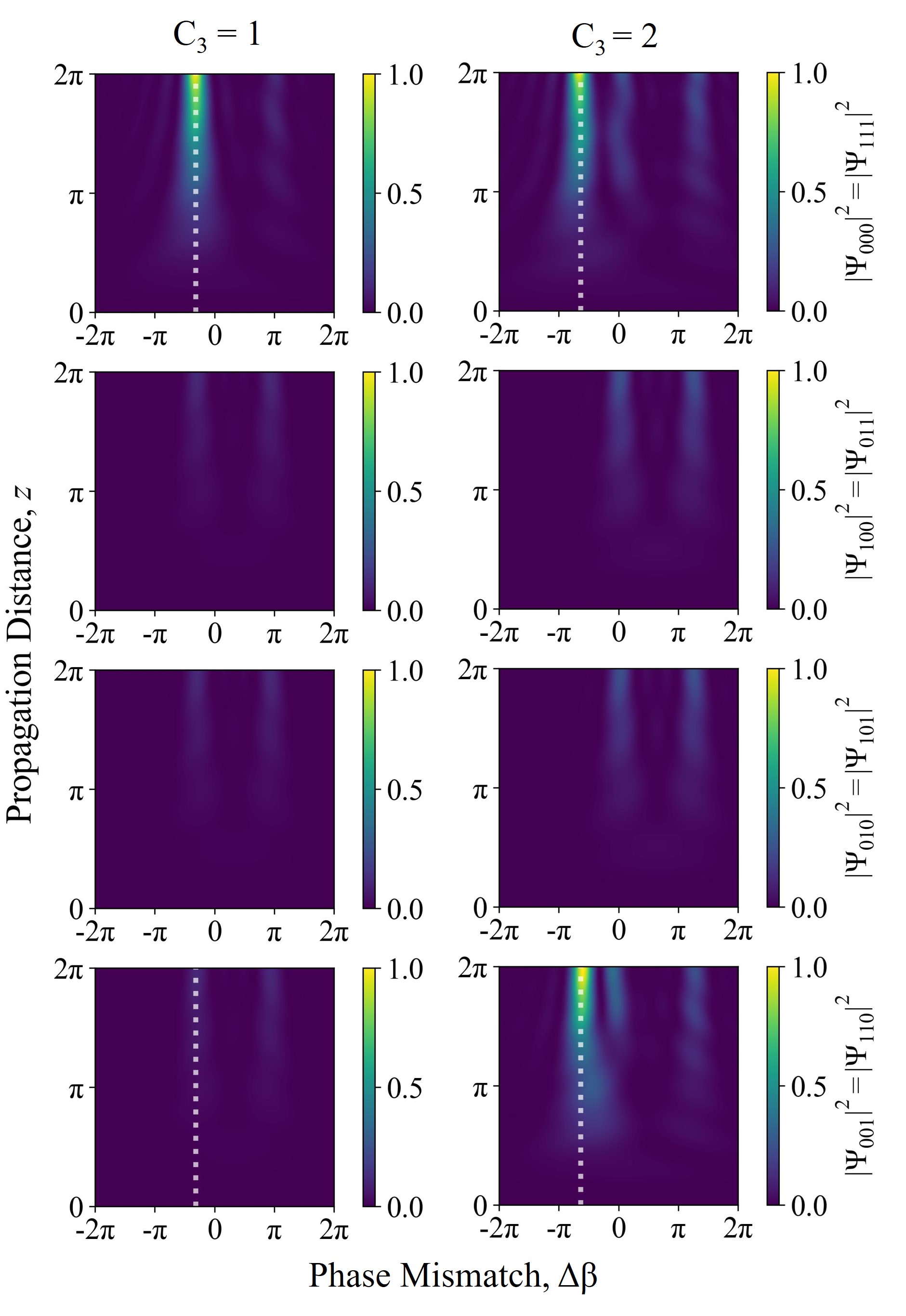}
\caption{Spatial dynamics of the three-photon wavefunction in real space as it propagates through the coupled waveguide system. The white dotted line indicates the location for $\Delta \beta=-C_3$.}
\label{fig:spatial_dynamics}
\end{figure}

Figure \ref{fig:spatial_dynamics} shows the numerically calculated spatial evolution of the wavefunctions for different phase mismatch $\Delta \beta$ at fixed $C_1=1$, for cases where $C_3=1$ and $C_3=2$. On the first column where the coupling coefficients are all equal, two distinct peaks at $\Delta \beta=-1$ and $\Delta \beta=3$ are observed for all states. However, only the states $\Psi_{\{000,111\}}$ exhibited relatively higher amplitude at $\Delta \beta=-1$ while the other states are observed to be equivalent and with significantly less amplitude. The origin of these peaks are attributed to the individual ansatz and occur when $\lambda_j=0$ where $\Phi_j(z)$ is just a linear function in $z$. Examining Equation \ref{eq:eigenvalues} for the current case, the condition $\lambda_j =0$ is satisfied only at $\Delta \beta =\{-1,3\}$. Furthermore, constructive interference among the ansatz could give rise to nontrivial enhancement in probability amplitude at a particular peak, which for this case, was observed for $\Psi_{\{000,111\}}$ at $\Delta \beta=-1=-C_3$. 

On the other hand, when $C_3=2\neq C_1$, three distinct peaks can now be observed at $\Delta \beta =\{-2,0,4\}$ with three distinct solutions $\Psi_{\{000,111\}}$, $\Psi_{\{001,110\}}$, and $\Psi_{\{010,101,100,011\}}$. In this case, we now have two solutions that exhibit amplitude enhancement at $\Delta \beta=-2=-C_3$, namely $\Psi_{\{000,111\}}$ and $\Psi_{\{001,110\}}$, while the rest are greatly attenuated. By keeping $C_1=C_2$ for the rest of the study, we maintain the existence of the three distinct solutions due to the structure of the general solution. This peculiar dynamics between photon-triplet generation and mode coupling between the waveguides is harnessed in this study to systematically engineer a wide array of tunable output quantum state via optical switching without external post-processing. As we have seen, two completely distinct quantum states were generated by a simple switch in the coupling coefficient $C_3$, which can be experimentally implemented by introducing a non-degeneracy on the SPDC process and by spectral filtering for measurement. Tuning the phase mismatch parameter $\Delta \beta$ can also achieve amplitude enhancement on selected wavefunctions which can be implemented by controlling the transverse modes of the waveguides at the desired photon frequencies as well as by introducing a phase difference in the input pumps.

\subsection{Heralded Bell State}

Among the coveted multiphoton states are heralded Bell states (HBS) where at least one quantum state heralds an optical Bell pair. In the three photon case, this can be ideally constructed when $|\Psi_{000}|^2=|\Psi_{111}|^2=|\Psi_{110}|^2=|\Psi_{001}|^2\neq0$ and others are zero. Here, the states $\Psi_{000}$ and $\Psi_{111}$ carry the Bell pair information and is being heralded by the measurement of the third photon i.e. the measurement the states $\Psi_{110}$ or $\Psi_{001}$. In this use case, the HBS condition imposes that:

\begin{equation}
    \begin{aligned}
        \begin{cases}
        |a+b+c+d|^2=|-a-b+c+d|^2\neq0 \\
        |a-b-c+d|^2=|-a+b-c+d|^2=0.
        \end{cases}
    \end{aligned}
\end{equation}

\begin{figure}
\centering
\includegraphics[width=\linewidth]{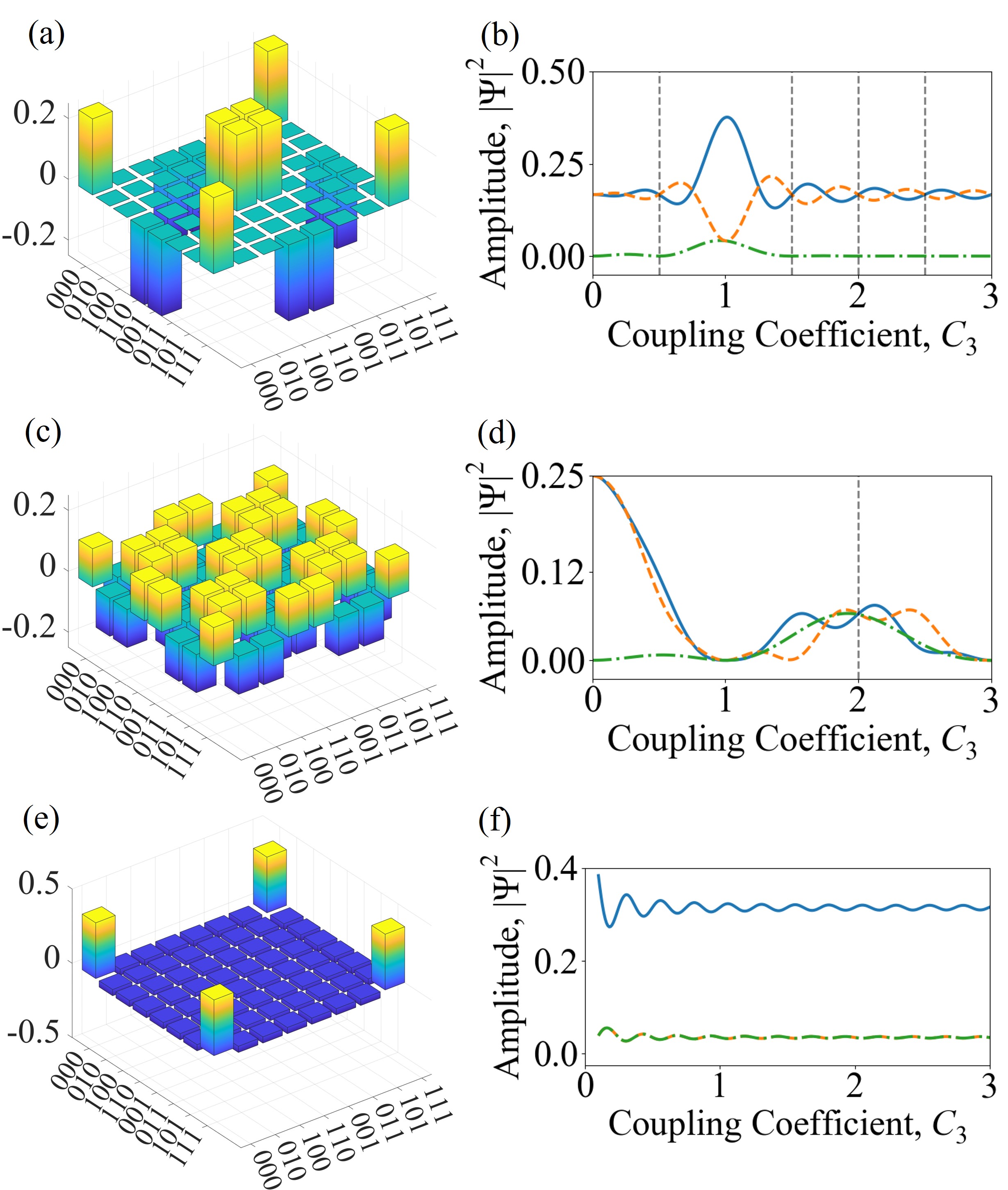}
\caption{ Real part of the calculated density matrices at the waveguide output ($z=2\pi$) for (a) heralded Bell State, (c) uniform state, and (e) GHZ-like state. Dependence to $C_3$ at fixed $C_1=C_2=1$ of the wavefunctions for (b) heralded Bell state, (d) uniform state, and (f) GHZ-like state. The blue solid lines, orange dashed lines, and green dash-dotted lines refer to three distinct solutions $\Psi_{\{000,111\}}$, $\Psi_{\{001,110\}}$, and $\Psi_{\{010,101,100,011\}}$, respectively.} 
\label{fig:density_matrix}
\end{figure}

This system of equations has several different solutions but within our parameter space i.e. $C_1,C_2,C_3>0$, it can be shown that the HBS can be constructed by satisfying the following condition: 

\begin{equation}
\begin{aligned}
&\begin{cases}
    C_3=-\Delta\beta=m/4, \\
    C_1+C_2=n/2,
    \end{cases}
\ \begin{cases}
    C_1=C_2=n/4, \\
    C_3-\Delta\beta=m/2, 
    \end{cases} \\ 
    &\begin{cases}
    C_3=-\Delta\beta=m/4,\\
    C_1=C_2=n/4,
    \end{cases}
\label{eqn:HBS_condition}
\end{aligned}
\end{equation}
where $\ m\neq n$ and  $n,m\in \mathbb{Z}\setminus\{0\}$. As an example, we recall our results from Figure \ref{fig:spatial_dynamics} for the case where $C_3=2$ and choose $\Delta \beta = -C_3$ (white dotted line) which clearly satisfies Equation \ref{eqn:HBS_condition}. At the waveguide output, the wavefunction can be described by the density matrix shown in Figure \ref{fig:density_matrix}a. Here, only the real part is displayed as the imaginary components are much smaller in magnitude. The condition specified above applies to creating a perfect HBS where the magnitude of the Bell pair carrier and the herald are equal while all other states are exactly zero. However, in  practice, finite noise exists that can cause the quantum state to diverge from its ideal form. For the case of HBS, this can manifest as an inequality in the finite amplitude of the Bell pair carrier and the herald, as well as by having infinitesimally small but non-zero amplitude for other states. As seen on Figure \ref{fig:density_matrix}b, this physically realistic description of the HBS is generally satisfied when $\Delta \beta=-C_3$ for all other $C_3$ values except those close to $C_1=C_2=1$ (See Supporting Information). Here, there is still finite amplitude for the Bell pair carrier and herald states while the rest are greatly attenuated. The dotted lines indicate the locations for perfect HBS where the unwanted wavefunctions are exactly zero. Thus, our system has demonstrated that we can effectively generate a robust HBS on a nonlinear coupled waveguide system by tuning the phase mismatch $\Delta \beta$ to $-C_3$ for any $C_3$ not close to $C_1=C_2$. 

Additionally, we note that the reverse HBS can also be constructed in the same principle where $|\Psi_{011}|^2=|\Psi_{100}|^2=|\Psi_{010}|^2=|\Psi_{101}|^2\neq0$ and others are zero. 

\subsection{Uniform State}

Moreover, uniform states, i.e. states where all basis wavefunctions have equal amplitude, are shown to be realizable with our proposed system. Such states require that $|\Psi_{l,m,n}|^2 = |\Psi_{l',m',n'}|^2$ for any numbers $l,m,n,l',m',n' \in \{0,1\}$. It can be created by setting $\Delta\beta=0$ and satisfying the following condition:

\begin{equation}
    \begin{cases}
    C_1=n/2,\\
    C_2=m/2, \\
    C_3= (n+m)/2
    \end{cases}
\end{equation}
where $n,m$ are any non-negative integers. Figure \ref{fig:density_matrix}c shows the density matrix for a uniform state constructed from the same parameters $C_1=C_2=1, C_3=2$ and by setting $\Delta \beta=0$. For this case, there is a strict requirement for $C_3$ to be the sum of $C_1$ and $C_2$. However, as seen on Figure \ref{fig:density_matrix}d, slight deviation to this strict $C_3$ requirement can still produce similar states but with small amplitude differences. Furthermore, we note that the generated uniform state, in contrast to the two-photon case, cannot be represented as completely separable since the three photon modes cannot have a unique spatial basis in the two coupled-waveguide system. This analytic observation is in congruence with recent findings where the three-mode correlation from TOSPDC was found to be fully inseparable in the general case.~\cite{Agusti2020}

\subsection{GHZ-like State}

Earlier, we have discussed the trivial case where the waveguides are uncoupled and produce a GHZ state. However, with finite coupling, it can be shown that it is still possible to make the waveguides effectively uncoupled and prepare GHZ-like states. To prepare such states, we maximize $\Psi_{\{000,111\}}$ and minimize all others. Such condition is shown to be satisfied when $\Delta \beta = -C_3$ and all coupling coefficients are equal, i.e. $C_1=C_2=C_3$. Figure \ref{fig:density_matrix}e shows the calculated density matrix, resembling that of GHZ state in the presence of finite noise in real-world scenarios. As can be seen in Figure \ref{fig:density_matrix}f, the generated GHZ state is preserved over a wide range of C-values.

\section{Conclusion and Outlook}

We have theoretically presented an approach to simultaneously generate and manipulate photon triplet states on an integrated nonlinear chip with optical tunability and no post-selection. The generation of photon triplet states is based on third-order SPDC in two coupled nonlinear waveguides excited by two pump beams. The built-in tunability is made possible by adjusting the phase mismatch $\Delta \beta$ and coupling coefficients $C_{\{1,2,3\}}$ which then predicts the nontrivial propagation dynamics of the generated photons. Furthermore, our proposed device was shown to be capable of producing robust heralded Bell states, uniform states, and GHZ-like states by appropriate parameter tuning. 

With the advent of on-chip integrated pump lasers, ~\cite{Wang2023,Yang2024} quantum photonic inter-converters, ~\cite{Wang2016,Feng2016} electro-optic modulators,~\cite{Rahim2021} and single-photon devices, \cite{Elshaari2020,Wang2022} it is now becoming possible to implement all major components of photonic quantum circuitry on a chip.  Hence, our work imparts another degree of flexibility for manipulating photonic quantum information for emerging advanced applications.

Additionally, due to the cubic nonlinearity of the waveguide, other $\chi^{(3)}$ nonlinear processes are allowed to exist such as third-harmonic generation (THG), spontaneous four-wave mixing (SFWM), self-phase modulation (SPM), and cross-phase modulation (XPM). \cite{Agrawal2014-ks,Lin2007} However, for THG, which shares the same phase matching condition as TOSPDC, the output photons are spectrally separated from the pump ($3\omega_p = \omega_{THG}$), and thus do not effectively propagate along the waveguide and interfere with TOSPDC photons. \cite{Boyd2008,Moebius2016} On the other hand, the different phase-matching requirement for SFWM restricts its efficiency which is often not satisfied in the TOSPDC process. \cite{Boyd2008,Solntsev2012_cubicSFWM} Self-phase modulation due to intensity-dependent refractive index (optical Kerr effect) introduces a phase shift to the pump resulting to an effective propagation constant of form: $\tilde{\beta_{p}} = \beta_{0 p} + \gamma_{SPM} P_p $, where $\beta_{0p}$ is the unperturbed propagation constant, $\gamma_{SPM}$ is the SPM strength, and $P_p$ is the pump power.  \cite{Banic2022} Cross-phase modulation of the TOSPDC photons due to the pump similarly results to an effective propagation constant $\tilde{\beta_{i}} = \beta_{0i} + \gamma_{XPM} P_p$. \cite{Banic2022} It will be an interesting subject of future work to explicitly include the contribution of these nonlinearities and losses in the model. The existence of $\chi^{(2)}$ or pseudo-$\chi^{(2)}$ wave mixing processes in centrosymmetric $\chi^{(3)}$ materials has also been recently observed and investigated, however, sophisticated bandstructure engineering would be required to induce such effects and may become the basis for future work. \cite{Rodriguez2007,Cazzanelli2011,Castellan2019,Soref2023}

 Moreover, incorporating resonant components such as ring resonators \cite{Banic2022} and metasurfaces \cite{Koshelev2019,Marino2019,SantiagoCruz2021} may offer improved spectral control of the three-photon generation as well as a boost in the overall efficiency, which currently remains low and is an active subject of ongoing research. 

Lastly, the integration of our proposed system in generating hyperentangled states is another promising direction where the generation of entangled photons and the encoding of its spatial degree of freedom are incorporated in the waveguide. This has the potential to reduce the number of optical elements needed to generate multiphoton hyperentangled states with tripartite entanglement similar to the protocol recently discussed in the literature. \cite{Zhao2023,Zhao2024} We anticipate the experimental adoption and verification of our theoretical analysis to find its relevance in path-encoded quantum information.

\medskip
\textbf{Supporting Information} \par 
Supporting Information is available in the appendix.

\medskip
\textbf{Acknowledgements} \par 
The authors were supported by the Sydney Quantum Academy, Sydney, NSW, Australia. M.B. also acknowledges UTS for the support through the International Research Scholarship. 

\textbf{Disclosures} \par

The authors declare no conflicts of interest.

\textbf{Data availability} \par 

No data were generated or analyzed in the presented research.

\medskip

\bibliography{references}

\newpage
\appendix
\onecolumngrid
\textbf{Supporting Information}

\section{Analytical solution of the system dynamics}
As it was mentioned in the main article the system dynamics can be described by the system of equations on wavefunctions over coordinate $z$.

\begin{align}
\begin{split}
-i\frac{d}{dz}\Psi_{l,m,n} &= C_1 \left[ \Psi_{l-1,m,n} + \Psi_{l+1,m,n} \right]\\ 
&+ C_2 \left[ \Psi_{l,m-1,n} + \Psi_{l,m+1,n} \right] \\ 
&+ C_3 \left[ \Psi_{l,m,n-1} + \Psi_{l,m,n+1} \right]  \\&- i \sum_{p=0,1}  \delta_{m,p} \delta_{n,p} \delta_{l,p}  \gamma A_{p}  e^{i\Delta \beta z}.
\end{split}
\label{eqn:spatial_evolution}
\end{align}

By making a substitution $\Psi_{l,m,n}\rightarrow\Psi_{l,m,n}e^{i\Delta\beta z}$ we can represent a new inhomogeneous system of linear differential equations as
\begin{equation}
    \frac{d\Psi}{dz}=i\hat{A}\Psi+B,
\end{equation}
where, $\Psi=\left(\Psi_{000},\Psi_{100},\Psi_{010},\Psi_{001},\Psi_{110},\Psi_{101},\Psi_{011},\Psi_{111}\right)^{T}$, $B=\gamma\left(A_0,0,0,0,0,0,0,A_1\right)^{T}$ and

\begin{equation}
    \hat{A}=\begin{pmatrix}
-\Delta\beta & C_1 & C_2 & C_3 & 0 & 0 & 0 & 0\\
C_1 & -\Delta\beta & 0 & 0 & C_2 & C_3 & 0 & 0\\
C_2 & 0 & -\Delta\beta & 0 & C_1 & 0 & C_3 & 0\\
C_3 & 0 & 0 & -\Delta\beta & 0 & C_1 & C_2 & 0\\
0 & C_2 & C_1 & 0 & -\Delta\beta & 0 & 0 & C_3\\
0 & C_3 & 0 & C_1 & 0 & -\Delta\beta & 0 & C_2\\
0 & 0 & C_3 & C_2 & 0 & 0 & -\Delta\beta & C_1\\
0 & 0 & 0 & 0 & C_3 & C_2 & C_1 & -\Delta\beta
\end{pmatrix}
\end{equation}

By introducing the unitary transformation such as $\hat{D}=U\hat{A}U^{-1}$, where $\hat{D}=\text{diag}\{\lambda_j\}_{j=1,..,8}$ we can find that for $\Phi=U\Psi$ and $B'=UB$ 
\begin{equation}
    \Phi_j(z)=-\frac{1-e^{i\lambda_j z}}{i\lambda_j}B'_j.
\end{equation}
Note that if $\lambda_j=0$, then $\Phi_j=B'_jz$. From here it is clear that in general the solution for different $\Psi_{n,m,l}$ is a combination of oscillating and linear functions over $z$. Let's introduce some notations: $\lambda_a=-\Delta\beta +C_1-C_2-C_3$, $\lambda_b=-\Delta\beta -C_1+C_2-C_3$, $\lambda_c=-\Delta\beta -C_1-C_2+C_3$, $\lambda_d=-\Delta\beta +C_1+C_2+C_3$ and $\overline{\lambda}=\frac{1}{2}\sum\limits_{j=a,b,c,d}\lambda_j$. Then the eigenvalues of matrix $\hat{A}$ are as follows:
    \begin{equation}
    \hat{D}=\text{diag}\left\{
\overline{\lambda}-\lambda_d, \ \lambda_a, \ \lambda_b,\ \overline{\lambda}-\lambda_c,\ \lambda_c, \ \overline{\lambda}-\lambda_b, \ \overline{\lambda}-\lambda_a, \ \lambda_d
\right\}
\end{equation}
From the structure of the matrix $\hat{A}$ it is clear that only maximum 4 eigenvalues can be linear independent when others are not. 

From this moment we will assume that two waveguides have the same amplitude generation $\gamma A_1=\gamma A_0=\gamma'$. This fact extremely simplifies the general solution and if $\Phi=\left(0,a,b,0,c,0,0,d\right)^{T}$, where $a,b,c,d=\Phi_2(z),\Phi_3(z),\Phi_5(z),\Phi_8(z)$ respectively, then the initial wavefucntions can be represented as 
\begin{equation}
\begin{aligned}\label{eq:sol_gen}
    &\Psi_{000}=\Psi_{111}=\frac{1}{2\sqrt{2}}(a+b+c+d), \\ 
    &\Psi_{100}=\Psi_{011}=\frac{1}{2\sqrt{2}}(a-b-c+d), \\
    &\Psi_{010}=\Psi_{101}=\frac{1}{2\sqrt{2}}(-a+b-c+d), \\
    &\Psi_{001}=\Psi_{110}=\frac{1}{2\sqrt{2}}(-a-b+c+d).
\end{aligned}
\end{equation}
Such equality is a consequence of the symmetry that system achieves with respect to the permutation of waveguides. Further, we are going to find out what interesting states can be built by varying the initial parameters of the system with to essential restrictions, namely we are going to study states at the end of waveguides, i.e. we fix $L=z_{max}=2\pi$ and we assume the experimental requirement, i.e. $C_1,C_2,C_3>0$. 

\subsection{Heralded Bell state (HBS)}

From the solution ~\eqref{eq:sol_gen} the ideal HBS can be built having the structure $|\Psi_{000}|^2=|\Psi_{111}|^2=|\Psi_{110}|^2=|\Psi_{001}|^2\neq0$ and others are zero. Hence, the HBS condition for the $z=L$ can be rewritten as
\begin{equation}
    \begin{aligned}
        \begin{cases}
        |a+b+c+d|^2=|-a-b+c+d|^2\neq0 \\
        |a-b-c+d|^2=|-a+b-c+d|^2=0
        \end{cases}
    \end{aligned}
\end{equation}
This system has several different solutions but we will confine ourselves only to those are physical realizable, i.e. that preserves the fact that $C_1,C_2,C_3>0$. It effectively requires such condition on eigenvalues $\lambda_j$:
\begin{equation}
    \begin{cases}
        \lambda_c,\lambda_d\in \mathbb{Z}\setminus\{0\}\\
        \lambda_a,\lambda_b\notin \mathbb{Z}\setminus\{0\}
        \end{cases}
\end{equation}
The reason for confining some numbers to integers except zero is to gurrantee the respective $\Phi_j$ at the end of waveguides will be zero.
As a result we obtain 3 different solutions that 
\begin{equation}
\begin{aligned}
\begin{cases}
    C_3=-\Delta\beta=m/4, \\
    C_1+C_2=n/2,;
    \end{cases} \ 
\begin{cases}
    C_1=C_2=n/4,  \\
    C_3-\Delta\beta=m/2, 
    \end{cases} \ \begin{cases}
    C_3=-\Delta\beta=m/4,\\
    C_1=C_2=n/4,
    \end{cases}
\end{aligned}
\end{equation}
where $\ m\neq n$ and  $n,m\in \mathbb{Z}\setminus\{0\}$.

We should note that the same procedure can be applied to build a mirrored HBS where $|\Psi_{011}|^2=|\Psi_{100}|^2=|\Psi_{010}|^2=|\Psi_{101}|^2=0$ and others are zero.

\subsection{Uniform states}
Moreover uniform states can be constructed from the general solution ~\eqref{eq:sol_gen}. It requires $|\Psi_{l_1,m_1,n_1}|^2=|\Psi_{l_2,m_2,n_2}|^2$ for any numbers $l_{1,2},m_{1,2},n_{1,2}$. To find such state we put $\Delta\beta=0$ and we obtain the physical solutions as
\begin{equation}
    \begin{cases}
        \lambda_a,\lambda_b,\lambda_d\in \mathbb{Z}\setminus\{0\}\\
        \lambda_c\notin \mathbb{Z}\setminus\{0\}
        \end{cases}
\end{equation}
After some algebra it can be rewritten as
\begin{equation}
    \begin{cases}
    C_1=n/2,\\
    C_2=m/2, \\
    C_3= (n+m)/2
    \end{cases}
\end{equation}
where $n,m$ are any non-negative integers. Nevertheless, this solution cannot be represented as a separable state because of the symmetry of the system.

\subsection{Weak coupling}
It is obvious that when waveguides are uncoupled, all three photons can be only in the first or in the second waveguide, hence $|\Psi_{000}|^2=|\Psi_{111}|^2=1/2$ and others are zero. However, this regime is unique, because there is no finite $C_1,C_2,C_3,\Delta\beta$ that effectively make waveguides uncoupled, that can be easily derived from the solution of linear equations that has only trivial solution. Nevertheless, if coupling is weak, namely $C_1,C_2,C_3\ll1$ then the error of uncoupling can be estimated easily. 

The regime of weak coupling obtains when $\lambda_{a,b,c,d}\ll1$. Then by expanding over $z$ dependence wavefunctions we can get
\begin{equation}
    \frac{|\Psi_{100}|^2}{|\Psi_{000}|^2}=\frac{|C_1|^2z^2}{4+\Delta\beta^2z^2}, \ \frac{|\Psi_{010}|^2}{|\Psi_{000}|^2}=\frac{|C_2|^2z^2}{4+\Delta\beta^2z^2}, \ \frac{|\Psi_{001}|^2}{|\Psi_{000}|^2}=\frac{|C_3|^2z^2}{4+\Delta\beta^2z^2}
\end{equation}
It is obvious that the error increases with the distance of photons flying in waveguides.
To minimise these expressions at the end of waveguides we can assume (for example) $C_1=C_2=0$ and $C_3=|\delta\beta|$.  

\section{Heralded Bell State Solution ($\Delta \beta = -C_3$)}

\begin{figure}[h!]
    \centering
    \includegraphics[width=0.5\linewidth]{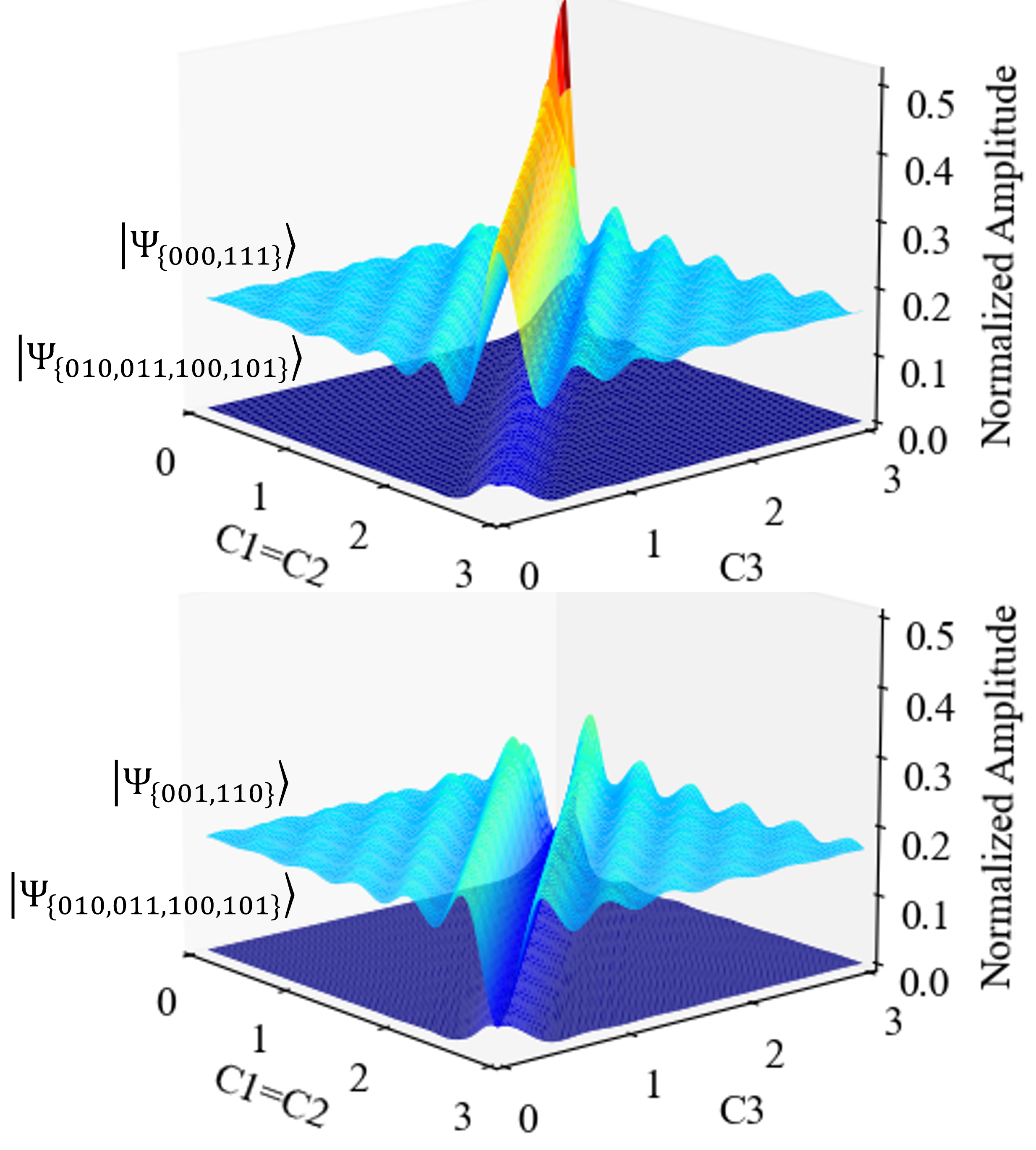}
    \caption{Amplitude of the states $\Psi_{\{000,111 \}}$ (top) and $\Psi_{\{001,110 \}}$ (bottom), each compared to the states $\Psi_{\{010,011,100,101 \}}$. }
    \label{sup_fig:HBS_C1}
\end{figure}

Figure \ref{sup_fig:HBS_C1} shows the calculated solution for the heralded Bell state condition at different $C_1=C_2$ and $C_3$. Here, it can be seen that the states $\Psi_{\{000,111 \}}$ and the $\Psi_{\{001,110 \}}$ have finite normalized amplitude $|\Psi_{l,m,n}|^2$ while the rest of the states are highty attenuated, except at $C_3$ close to $C_1$.

\end{document}

%% file: preamble.tex
\usepackage{amsthm}
\usepackage{mathtools}
\usepackage{physics}
\usepackage{xcolor}
\usepackage{graphicx}
\usepackage[left=23mm,right=13mm,top=35mm,columnsep=15pt]{geometry} 
\usepackage{adjustbox}
\usepackage{placeins}
\usepackage[T1]{fontenc}
\usepackage{lipsum}
\usepackage{csquotes}